# Three-dimensional structural and compositional inhomogeneity in zeolites unraveled by low-dose electron ptychography

Hui Zhang[1,2,3]*†, Guanxing Li[3]†, Jiaxing Zhang[4], Daliang Zhang[5], Zhen Chen[6], Xiaona Liu[7], Peng Guo[7], Yihan Zhu[8], Cailing Chen[3], Lingmei Liu[5], Xinwen Guo[4], and Yu Han[3]*

[1]Electron Microscopy Center, South China University of Technology, Guangzhou 510640, China

[2]School of Emergent Soft Matter, South China University of Technology, Guangzhou 510640, China

[3]Advanced Membranes and Porous Materials Center, Physical Science and Engineering Division, King Abdullah University of Science and Technology (KAUST), Thuwal 23955-6900, Saudi Arabia

[4]State Key Laboratory of Fine Chemicals, Frontiers Science Center for Smart Materials, School of Chemical Engineering, Dalian University of Technology, Dalian 116024, China

[5]Multi-scale Porous Materials Center, Institute of Advanced Interdisciplinary Studies & School of Chemistry and Chemical Engineering, Chongqing University, Chongqing 400044, China

[6]School of Materials Science and Engineering, Tsinghua University, Beijing 100084, China

[7]National Engineering Research Center of Lower-Carbon Catalysis Technology, Dalian National Laboratory for Clean Energy, Dalian Institute of Chemical Physics, Chinese Academy of Sciences, Dalian 116023, China

[8]Center for Electron Microscopy, State Key Laboratory Breeding Base of Green Chemistry Synthesis Technology and College of Chemical Engineering, Zhejiang University of Technology, Hangzhou 310014, China

*Corresponding authors. Email: hui.materials.zhang@gmail.com **(**H.Z.**)**; yu.han@kaust.edu.sa **(**Y.H.**)**

†These authors contributed equally to this work.

**Abstract**

Structural and compositional inhomogeneity is common in zeolites and considerably affects their properties. Conventional transmission electron microscopy (TEM) cannot provide sufficient information on local structures in zeolites due to the thickness-limited lateral resolution, lack of depth resolution, and electron dose-constrained focusing. We demonstrate that a multislice ptychography method based on four-dimensional scanning TEM (4D-STEM) data can overcome these limitations. The 4D-STEM ptychography image obtained from a ~40-nm thick MFI zeolite exhibits an ultrahigh lateral resolution of ~0.85 Å, enabling the unambiguous identification of individual framework oxygen (O) atoms and the precise determination of the orientations of adsorbed molecules. Furthermore, a depth resolution of ~6.6 nm is achieved, allowing the three-dimensional probing of O vacancies and phase boundaries in zeolites. The results of the 4D-STEM ptychography provide an unprecedented complete understanding of the intergrowth of MFI and MEL zeolites in three dimensions. The 4D-STEM ptychography can be generally applied to various zeolites and other materials with similar electron beam sensitivity.

**INTRODUCTION**

Zeolites are inorganic microporous crystals with regular intra-crystalline cavities and channels of molecular dimensions. The unique molecular sieving ability of zeolites renders highly desirable size- and shape-selectivity for adsorption, separation, and catalysis[1–4]. However, the complexity of zeolite structures makes them prone to structural and compositional inhomogeneity that can profoundly affect their properties. For example, the precursors and synthetic conditions affect the location and distribution of functional sites (e.g., Al pairs and heteroatoms) in the zeolite framework[5]. Post-synthesis calcination generates oxygen (O) vacancies to render Lewis acid sites in an uncontrolled manner[6,7]. The intergrowth of multiple polymorphs and various phases is common in zeolite crystals[8–10]. The prevalence of inhomogeneity explains why zeolites with the same topology and composition prepared from different batches often behave differently in applications[11,12].

Although inhomogeneity in zeolites has been widely recognized, the precise determination of the related local nonperiodic structures remains challenging due to the lack of suitable characterization techniques. Transmission electron microscopy (TEM) is one of the most commonly used tools in materials science, capable of probing local structures at an atomic resolution through real-space imaging[13–15]. However, due to the electron beam sensitivity of zeolites, TEM imaging of zeolites must be performed under low electron dose conditions to prevent structural damage, resulting in images with dose-limited resolution and poor signal-to-noise ratio (SNR). These problems can be largely overcome by developing new imaging methods that use electron dose more efficiently[16,17].

Among emerging low-dose TEM techniques, integrated differential phase contrast scanning TEM (iDPC-STEM) has proven effective for imaging zeolite structures, especially for studying guest species located in the micropores[18–20]. Despite the demonstrated successes, iDPC-STEM has limitations. First, the image resolution of iDPC-STEM is severely limited by the specimen thickness. For zeolites, iDPC-STEM can only resolve framework T (T = Si or Al) atoms (strictly speaking, atomic columns) but not the O atoms between them unless the specimen thickness is within a few unit cells (<10 nm), which is extremely rare in actual samples. Second, like other conventional TEM imaging modes, iDPC-STEM lacks resolving power along the projection direction and cannot identify longitudinal structural inhomogeneity inside the specimen. Third, iDPC-STEM requires precise focusing of the electron beam to achieve high resolution, which could lead to a lower success rate for beam-sensitive materials because structural damage may occur during the fine-tuning of the focus.

Here, we demonstrate that electron ptychography[21–26] based on four-dimensional STEM (4D-STEM) data can solve these problems. Using 4D-STEM ptychography, we achieved a subangstrom resolution to resolve individual O atomic columns in various zeolites with specimen thicknesses up to ~40 nm, which is impossible with any existing direct-imaging technique. Clearly identifying O atoms enables a semi-quantitative analysis of the three-dimensional (3D) distribution of O vacancies throughout the zeolite specimen. Furthermore, 4D-STEM ptychography provides ~6.6 nm depth resolution to reveal the growth of MEL domains along the *b*-axis of the MFI zeolite, which was previously unknown. In addition, acquiring a 4D-STEM ptychography dataset does not require precise focusing of the electron probe, considerably improving the efficiency of obtaining atomic-resolution phase-contrast images of zeolites and other beam-sensitive materials.

**RESULTS AND DISCUSSION**

Image simulations were first performed to demonstrate the superiority of 4D-STEM ptychography over iDPC-STEM, the state-of-the-art technique for zeolite imaging. A series of 4D-STEM datasets were simulated for [010]-projected zeolite ZSM-5 by varying the specimen thicknesses and convergence angles (see EXPERIMENTAL SECTION below for details). Ptychography and iDPC-STEM images were calculated from each dataset. The results reveal that, regardless of the convergence angle, when the sample thickness exceeds 12 nm, iDPC-STEM images start to deviate from the structural model and become difficult to interpret. In contrast, ptychography images are very robust, displaying substantial tolerance for the specimen thickness variation of up to 40 nm (Fig. S1).

This remarkable advantage of ptychography was experimentally verified using specially synthesized ZSM-5 zeolite crystals. These ZSM-5 crystals have a highly uniform thickness of 30–40 nm along the *b*-axis, as evidenced by multiple characterizations (Fig. S2), allowing a reasonable comparison between iDPC-STEM and ptychography. The iDPC-STEM images were acquired using conditions optimized for zeolite imaging (convergence semi-angle: 15 mrad; probe current: 2 pA; dwell time: 10 μs; pixel size: $0.380 \times 0.380$ Å$^2$), with a total electron dose of ~900 e$^-$/Å$^2$. The 4D-STEM datasets were collected using a high-performance hybrid pixel array detector EMPAD with a probe current of ~0.09 pA, scan step of 0.399 Å, and 256 × 256 probe positions. Limited by the EMPAD frame rate (1000 frames per second), the specimen was subjected to an electron dose of ~3500 e⁻/Å² during data acquisition, which proved to be sufficiently low to preserve the structure of ZSM-5. Iterative ptychographic reconstructions were performed using the maximum likelihood (ML) and linear least squares (LSQ) methods to update the exit wave function and object and probe functions, respectively[27,28]. Multislice[29] and mixed-state[30] algorithms were implemented to address the multiple scattering associated with thick specimens and the partial incoherence of electron probes, respectively. Successful multislice mixed-state ptychographic reconstruction at low-dose conditions relies on the best possible initial guess for the probe function, which can be facilitated by minimizing aberrations using the clever data acquisition strategy described in Fig. S3.

The iDPC-STEM image taken from ~40-nm thick ZSM-5 does not match the [010]-projected structural model and is difficult to interpret (Fig. S4), which is consistent with the simulation. When the crystal periphery, which was significantly thinner than the bulk region, was chosen for imaging, 5-, 6-, and 10-membered rings surrounded by T atoms can be identified in the acquired iDPC-STEM image, but the O atoms between the T atoms remain unresolvable (Fig. 1a).

Clear identification of O in the zeolite framework requires a resolution close to 1 Å and good image contrast, which is rarely achievable with iDPC-STEM due to its sensitivity to sample thickness and time-constrained focusing unless the specimen is extremely thin[20]. Multislice ptychography effectively overcomes these limitations. Figure 1b illustrates the ptychographic phase image of ~40-nm thick ZSM-5, where all framework atoms, including O atoms, are unequivocally resolved. This image perfectly matches the projected structural model (Fig. 1c) and involves structural information transferred up to 0.85 Å (Fig. S5), corresponding to a record-high resolution for zeolite imaging using electron microscopy. The 4D-STEM data were collected at a high magnification after only coarse focusing at low magnification. Ptychography reconstruction revealed that the electron probe was ~35 nm defocused (Fig. S6), which could not provide atomic resolution using conventional STEM modes. Not having to fine-tune the beam focus is a considerable advantage of 4D-STEM ptychography, which is especially important for beam-sensitive materials.

Similar results were obtained when ZSM-5 was imaged along the [100] axis. Thin specimens in this orientation were obtained by crushing commercial micron-sized ZSM-5 crystals into small fragments. While the exact specimen thicknesses were unknown, we selected the highest quality iDPC-STEM and ptychographic images obtained for comparison (Fig. 1d and 1e). The results confirmed the significantly higher resolving power of 4D-STEM ptychography compared with iDPC-STEM. In particular, there are four closely spaced T atomic columns in this projection (T-T distances: 0.8~1.8 Å; Fig. 1f), which are merged together in the iDPC-STEM image (Fig. 1d) but are clearly separated in the ptychographic image (Fig. 1e). The separation of these four atomic columns has not previously been achieved by imaging. The Fourier transforms of the images in Fig. 1 are presented in Fig. S5 to quantitatively illustrate the difference in resolution between 4D-STEM ptychography and iDPC-STEM. When used for other types of zeolites, such as EMM-17, with unusual 11-MR channels, 4D-STEM ptychography also demonstrated sub-Å resolution and the ability to identify the framework O (Fig. S7).

The 4D-STEM ptychography was used to image ZSM-5 with adsorbed *p*-xylene (PX) molecules. The reconstructed ptychographic image shows rod-like and fluffy-dot-like contrasts in the straight 10-MR channels, which are generally considered to correspond to PX molecules in edge-on and face-on orientations, respectively (Fig. 1g). While previous studies have reported similar observations using iDPC-STEM[19,31], the ultrahigh spatial resolution of 4D-STEM ptychography allows for more precise determination of the molecular orientations. As shown in Fig. 1h, the PX molecules pointing to the T atom, O atom, and middle of the T–O bond were all identified. The diverse orientations of adsorbed molecules imply the presence of complex host–guest interactions in zeolites due to their inhomogeneous chemical environments.

Zeolites have Brönsted and Lewis acidity, both playing vital roles in catalysis. The O vacancies generated by framework dehydration during high-temperature calcination is generally considered to be a major source of Lewis acidity, whereas direct observation of O vacancies in zeolites has not been realized to date.

The unprecedented ability of 4D-STEM ptychography to image O in the zeolite framework enables the identification of O vacancies. To demonstrate this application, we collected 4D-STEM data from a ZSM-5 sample with a Si/Al ratio of ~100 along the [010] axis and reconstructed its phase image using the multislice ptychography method described above. The reconstruction result consists of seven slices, each with a thickness of 4 nm (i.e., 2 unit cells); thus, the imaged area is ~28 nm thick. Owing to the ultrahigh resolution of the ptychographic image, the T and O atomic columns could be precisely identified in each slice using an atomic image recognition software (Fig. S8) and subsequently used for intensity analysis.

The intensity distribution of T columns plotted from the experimental ptychographic image was consistent with the simulation based on the defect-free ZSM-5 structure, both displaying a 3% coefficient of variation (Fig. 2a). In contrast, the O columns in the experimental ptychographic image exhibited a more pronounced intensity fluctuation than the simulation result (coefficient of variation: 7% *vs*. 5%; Fig. 2b). Further analysis revealed that the substantial intensity fluctuation in the experimental image is due to a small fraction of the O columns that are significantly less intense than the rest (Fig. 2c), suggesting the presence of O vacancies in the actual sample.

Image simulations were performed using a series of ZSM-5 structural models with different O vacancy degrees to explore the effect of O vacancies on the intensity of O columns. The results indicate that one O vacancy within a thickness of 4 nm cannot be reliably determined because the intensity reduction it causes is within the intrinsic intensity fluctuation of ZSM-5. When two or more aligned O vacancies are within a thickness of 4 nm, the resulting intensity reduction is sufficient for a reliable assignment (Fig. 2c).

Based on the simulation results (Fig. 2c), we established a relative intensity threshold of 81% to identify O vacancies. We then analyzed the ptychographic image of the ~28-nm thick ZSM-5 slice by slice to render a 3D distribution of O vacancies within the crystal (Fig. 2d). The result revealed that the imaged region of the zeolite contained ~51 μmol O vacancies per gram, with 58% of these O vacancies exposed to the straight 10-MR channels (Table S2). It should be noted that the concentration of O vacancies obtained using this method needs to be treated with caution and preferably confirmed by other characterization techniques, as the result is sensitive to the choice of intensity threshold. For instance, if the relative intensity threshold of 80% is chosen, the concentration of O vacancies is ~24 μmol per gram.

Diffraction-based characterizations revealed that ZSM-5 and ZSM-11 often coexist in synthetic products due to their similarity in topology (MFI and MEL, respectively)[32,33]. However, it is challenging to determine the exact coexistence state, such as whether they are mixed or intergrown, and how the minor phase is distributed relative to the major phase. The only direct evidence for the intergrowth of MFI and MEL by electron microscopy was obtained from a 2D single-unit-cell thick zeolite membrane, in which the interconnection of two phases along the MFI *a*-axis with the MFI/MEL interface parallel to the *c*-axis was observed[34]. However, electron microscopy imaging has been unable to explore the 3D intergrowth of zeolite crystals due to the lack of resolving power in the depth direction.

We demonstrate that electron ptychography provides a solution to this problem using a zeolite material containing a primary phase, MFI, and concomitant phase, MEL, with an overall Si/Al ratio of ~90 (see EXPERIMENTAL SECTION below for the synthesis method). Comparison between experimental powder x-ray diffraction (PXRD) and PXRD simulation based on random stacking disorder[35] indicated that MEL constitutes 10% to 20% in the material (Fig. S11). The 4D-STEM ptychographic image of this material reveals subtle deviations from the typical structural projection along the MFI *b*-axis, where there appear to be some extra T atoms in the pentasil chain (Fig. 3a). The identification of extra T atoms indicates the presence of another structure besides the MFI in the projection direction. Based on the PXRD results, we speculated that the imaged zeolite crystal contained MFI and MEL structures along the MFI *b*-axis. The two structures of the pentasil family differ only in how the pentasil layers are connected (inversion symmetry *vs*. mirror symmetry), and their stacking along the *b*-axis results in a perfect match of the structure projection to the ptychographic image (Fig. 3b).

As the reconstructed ptychographic image consists of seven 4 nm-thick slices, a careful slice-by-slice analysis allowed us to confirm the intergrowth of MFI and MEL along the *b*-axis while probing their interfaces in the *a-c* plane at atomic resolution (Video S1). In the first two slices, a narrow MEL domain consisting of only two pentasil chains was observed, sandwiched by MFI domains (Fig. 3c). In the fourth slice, the MEL domain expanded, resulting in the generation of a new MFI/MEL interface within the pentasil chain (Fig. 3d). These observations confirm that MEL and MFI structures are intertwined in three directions within the crystal. Figures 3c and 3d clearly display how MFI and MEL structures are interconnected in the *a-c* plane. Along the *a*-axis, the two structures are perfectly interconnected by sharing a pentasil chain without generating strain or defects (Fig. 3c), which is consistent with the previous observation.[34]

The connection of the two structures along the *c*-axis (i.e., within the pentasil chain) can adopt three possible configurations (Fig. S13), of which the one with the least interfacial mismatch was observed (Fig. 3e). This is the first atomic-resolution observation of MFI/MEL intergrowth in this direction. Although the MFI/MEL interface perpendicular to the *b*-axis cannot be directly observed from ptychography due to limited depth resolution, it can be deduced based on the determined orientation relationship. When viewed along the *c*-axis, the interface displays alternating "connected" and "disconnected" regions along the *a*-axis, each with a width of one-half the unit cell parameter *a* (Figs. 3f and S14). The "disconnected" interface cannot accommodate additional $TO_4$ tetrahedra and should therefore be terminated by dangling silanol groups (Fig. 3g).

The depth resolution of 4D-STEM ptychography, which is affected by imaging conditions and reconstruction algorithms, must be determined explicitly for a given system. Using the 4D-STEM data acquired in this study and the LSQ-ML multislice algorithm, the minimum slice thickness for stable reconstruction is 4 nm, which defines the upper limit of depth resolution. The distinct structures of the two slices separated by 8 nm demonstrate a depth resolution better than 8 nm. To more precisely determine the depth resolution, we analyzed the intensity variation of two T atomic columns in seven consecutive slices spanning from one side of the MEL/MFI interface to the other, where the two studied columns were from MEL and MFI (Fig. 4a). In the interface region (i.e., Slices 3 and 4), the two columns interfere in intensity due to the limited depth resolution (Fig. 4a). The two intensity profiles were fitted with error functions and their derivative curves were then derived (Fig. 4b). The full width at half maximum of the derivative curve, which is generally defined as the depth resolution, was measured at ~6.6 nm (Fig. 4b). Using a 4D-STEM dataset simulated at a total electron dose of ~3500 $e^-/Å^2$, a better depth resolution of ~3.5 nm could be achieved based on the same determination method, due to the absence of experimental errors (Fig. S15).

We compared the results of 4D-STEM ptychography in this study with previous results obtained using various 3D imaging methods in terms of lateral resolution, depth resolution, and consumed electron dose (Fig. 4c). Electron tomography can provide an atomic resolution in three dimensions[36,37]. Optical sectioning with high-angle annular dark-field STEM can generate sub-nanometer depth resolution with dense focal series[38]. However, these two methods require very high electron doses, typically millions of $e^-/Å^2$, and are unsuitable for beam-sensitive materials (Fig. 4c, Table S3). Ptychographic tomography exhibited ~2 nm resolution in three dimensions at low-dose conditions (Fig. 4c)[39].

**CONCLUSIONS**

This study demonstrates that 4D-STEM multislice ptychography represents an efficient low-dose 3D imaging technique that exhibits excellent tolerance to the specimen thickness while not requiring precise focusing. These advantages make 4D-STEM multislice ptychography broadly applicable and particularly useful for imaging beam-sensitive materials. The currently achieved lateral and depth resolutions are ~0.85 Å and ~6.6 nm, respectively, using a total electron dose of ~3500 $e^-/Å^2$. This performance has enabled an unprecedented investigation of the intrinsic structural and compositional inhomogeneity in zeolites, including guest molecule orientations, O vacancies, and multiphase intergrowth. The simulations indicate that, under the employed imaging conditions, the depth resolution can, in principle, be improved to ~3.5 nm. We believe that combining multislice ptychography with tomography is worth exploring because of its potential to enhance 3D resolution further.

**EXPERIMENTAL SECTION**

The zeolite materials investigated in this work include ultrathin ZSM-5 (30–40 nm along the *b*-axis; Si/Al≈100), commercial ZSM-5 (Si/Al≈100~200), and ZSM-5/ZSM-11 (MFI/MEL; Si/Al≈90). Ultrathin ZSM-5 was synthesized using the previously reported protocol.[40] In this study, commercial ZSM-5 was purchased from Alfa Aesar in ammonium form, which was converted to the H-form through calcination at 600°C for 5 h before imaging.

Synthesis of ZSM-5/ZSM-11

For synthesis, 2.65 g of deionized water, 31.14 g of tetrapropylammonium hydroxide aqueous solution (25 wt.%), 40.00 g of silica solution (30 wt.% $SiO_2$ water suspension), and 3.53 g of silicalite-2 (seed suspension) were mixed and stirred at 35°C. In a separate beaker, 0.70 g of $Al_2(SO_4)_3·18H_2O$ and 1.11 g of $NH_4F$ were dissolved in 36.00 g of deionized water. The resulting solution was added to the mixture formed in the last step to make a gel. The gel was transferred into a 200 mL stainless-steel autoclave and heated at 170°C for 72 h. The solid products were filtered and washed with deionized water three times, dried at 80°C overnight, and then calcined at 550°C for 6 h.

Adsorption of *p*-xylene in ZSM-5

Ultrathin ZSM-5 was used to adsorb *p*-xylene. Specifically, 20 mg of freshly calcined ultrathin ZSM-5 was mixed with 1 mL of *p*-xylene, and the mixture was ultrasonicated for 3 h. Then, the mixture was dried in an oven at 100°C for 4 h.

4D-STEM data and iDPC-STEM image acquisition

The 4D-STEM and iDPC-STEM experiments were performed at 300 kV on an FEI Themis

Z microscope equipped with two aberration correctors. The convergence semi-angle ($α$) was 15 mrad. The preparatory steps, including the sample search, zone axis alignment, and aberration correction, are described in Fig. S3 with schematic illustrations.

The 4D-STEM datasets were collected using an EMPAD detector with 128 × 128 pixels.[41] The camera length was chosen to ensure that the bright-field disk occupies roughly half of the detector. The probe current used for 4D-STEM was set to ~0.09 pA, which can be measured directly by EMPAD. The raster scanning step and exposure time per frame were 0.399 × 0.399 Å$^2$ and 1.0 ms, respectively, corresponding to a total electron dose of ~3500 e$^-$/Å$^2$. The iDPC images were collected with a probe current of ~2.0 pA. The pixel size and dwell time were 0.380 × 0.380 Å$^2$ and 10 μs, respectively. The total electron dose of each iDPC image was calculated at ~900 e$^-$/Å$^2$.

Ptychographic reconstructions and data analysis

The multislice and mixed-state LSQ-ML algorithm[27] was implemented in MATLAB for ptychographic reconstructions, using a slice thickness of 4 nm. The tests revealed that a slice thickness smaller than 4 nm fails in reconstruction from the low-dose datasets. The sample thickness was determined by finding the appropriate number of slices that results in the maximum phase in the reconstructed phase images. The regularization factor, by which the information of the object function of different slices at low spatial frequencies in Fourier space was weighted,[22] was set to 1 for the initial sample thickness determination and 0.1 for the fine reconstruction.

Atomap[42] was used to analyze the atomic columns. The positions of the O and T columns were accurately determined by fitting each column with Gaussian functions. Subsequently, the intensity of each column was calculated by integrating the intensity of all pixels within a radius of 0.4 $x$~0.5 $x$, where $x$ is the distance between the column to be analyzed and its nearest neighboring columns.

Simulation of 4D-STEM data

The 4D-STEM datasets were simulated with abTEM[43] using Kirkland's parametrization of atomic potentials. The aberrations were set to zero for simplicity. Thirty frozen phonon configurations were averaged to account for thermal diffuse scattering. The standard deviations for the random thermal motions of Si and O were 0.075 and 0.1 Å, respectively. Datasets employed in Fig. S1 have infinite electron doses. For the simulated datasets in Figs. 2 and S15 and Table S1, the Poisson noise was added to simulate the effect of an electron dose of ~3500 e$^-$/Å$^2$. The 4D-STEM datasets were Gaussian-blurred to account for a finite source size of ~1 Å. The iDPC-STEM and ptychographic phase images were calculated and reconstructed from the simulated 4D-STEM datasets.

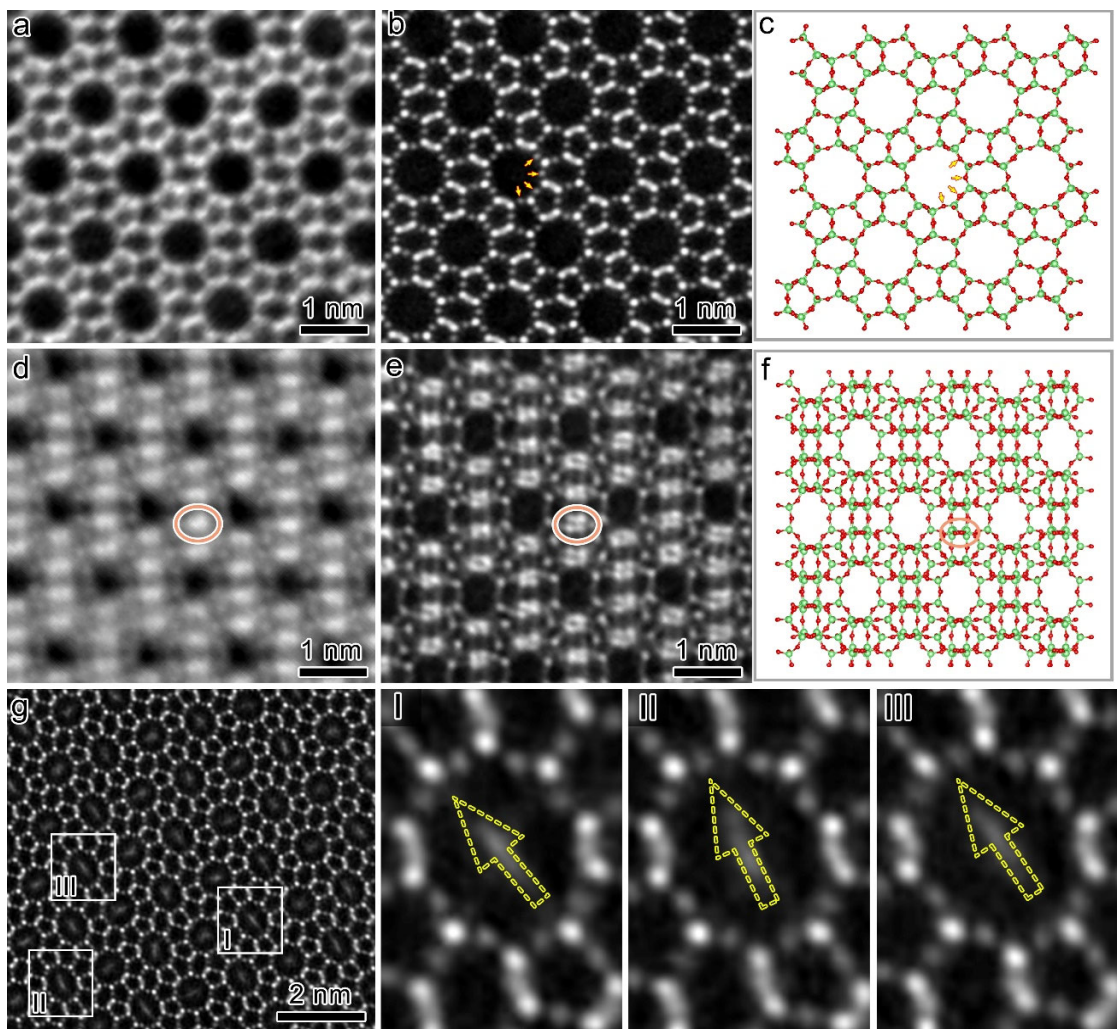

**Fig. 1** (a) iDPC-STEM image acquired from the periphery of a ZSM-5 crystal along the [010] axis, where the crystal thickness is estimated to be < 20 nm. (b) 4D-STEM ptychography image acquired from the bulk region of a 40-nm-thick ZSM-5 crystal along the [010] axis. (c) The [010]-projected structural model of ZSM-5. Arrows in (b) and (c) indicate several O atoms (atomic columns) in the ZSM-5 framework, which are not resolved in (a). (d) iDPC-STEM image of ZSM-5 along the [100] axis. (e) 4D-STEM ptychography image of ZSM-5 along the [100] axis. (f) The [100]-projected structural model of ZSM-5. Orange ellipses in (d-f) highlight the same positions, where four closely spaced columns of T atoms are resolved in (e) but not in (d). (g) 4D-STEM ptychographic image of ZSM-5 with adsorbed p-xylene (PX) molecules along the straight 10-MR channel (i.e., the [010] axis). Square regions I, II, and III are enlarged to reveal the subtly different orientations of PX molecules, as reflected by the rod-like contrast in the channels. Arrows are used as visual guides.

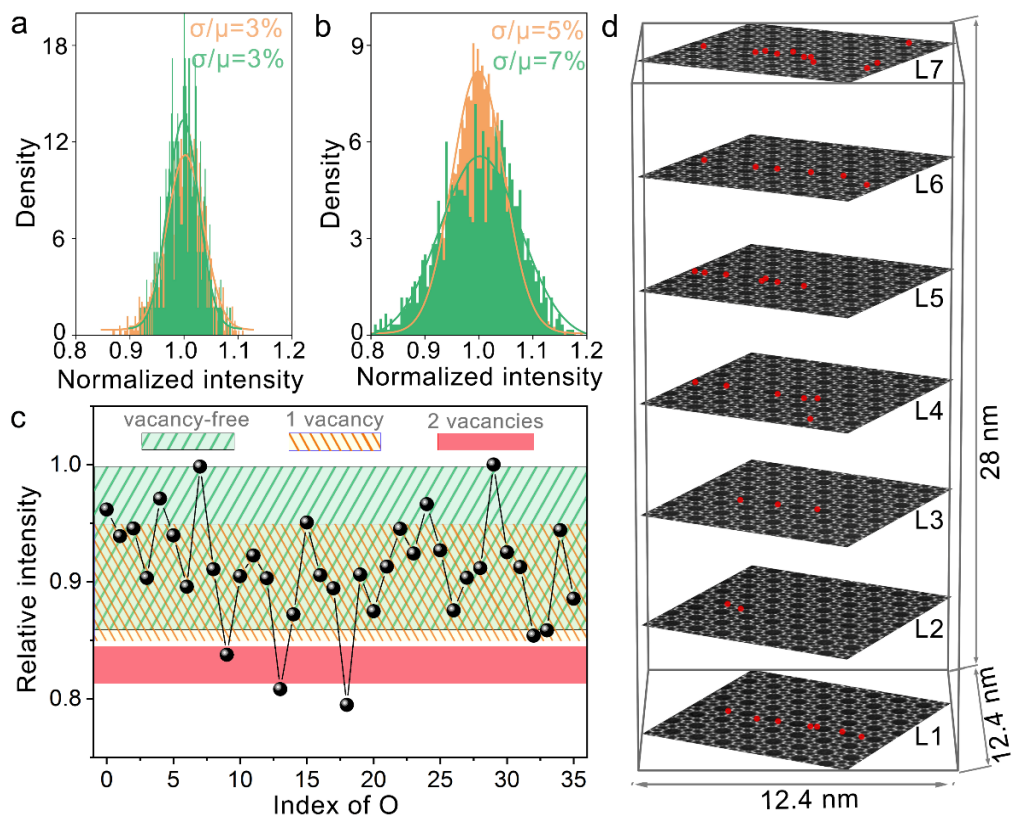

**Fig. 2** (a-b) Intensity histograms of (a) T columns and (b) O columns in the experimental (green) and simulated (orange) ptychographic phase images; σ is the standard deviation, and μ is the mean. The intensities are normalized by μ. The statistics are based on more than 1000 atoms. (c) Intensity variation of 30 O columns in the experimental ptychographic images. The image with atomic index is presented in Fig. S9. The intensity of each column is normalized by the maximum intensity in the 1 nm range. The light green region represents the intensity range of the O columns in the simulated ptychographic image using a vacancy-free model. Light yellow and red regions mark the intensity ranges of O columns containing one and two O vacancies per eight O sites, respectively. The model used for simulation is presented in Fig. S10, and the intensity data are listed in Table S1. In experimental images, the columns with an intensity smaller than the lower bound of the red region are identified as vacancy-containing columns. (d) Distribution of O vacancies identified by this method throughout the investigated region.

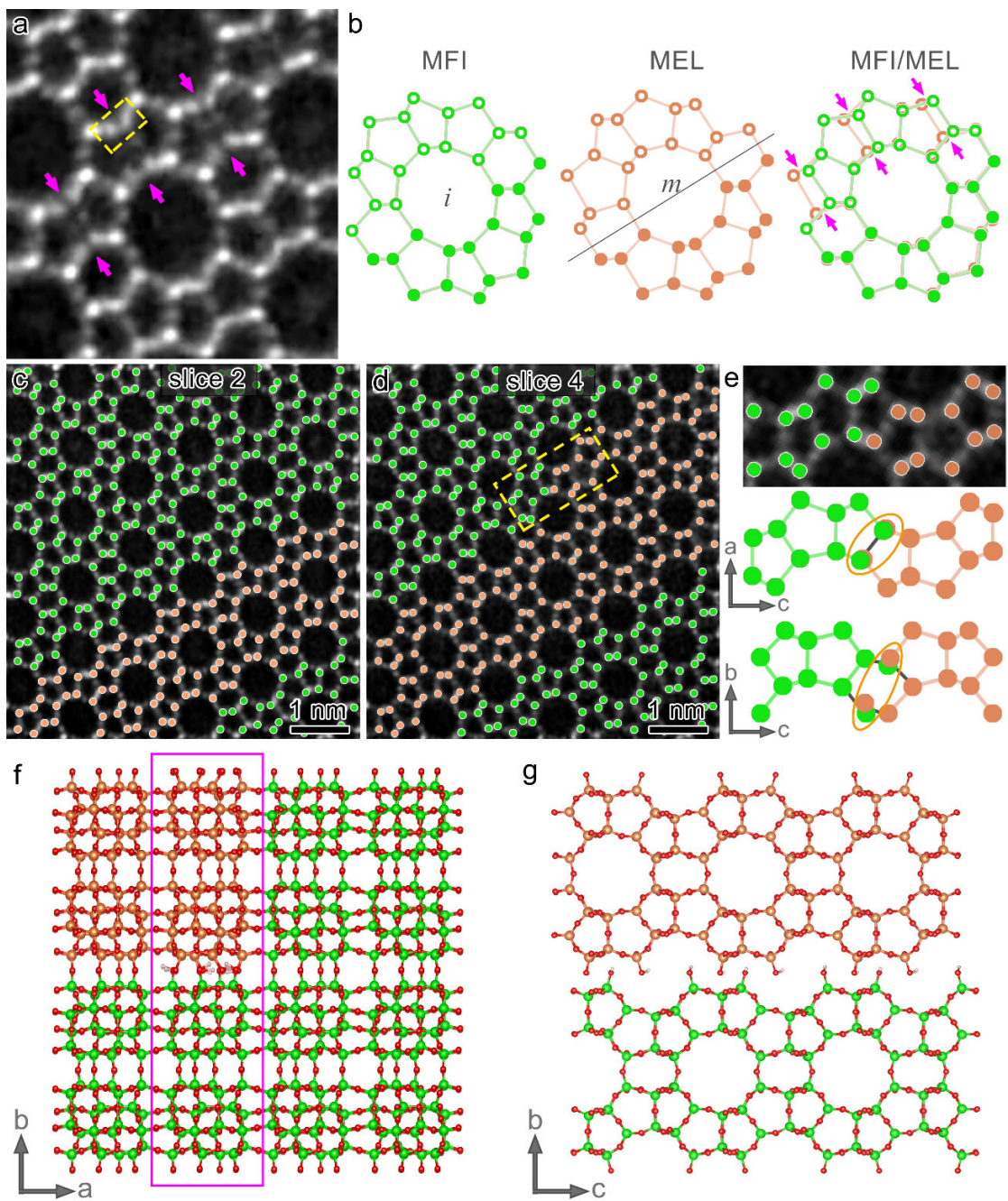

**Fig. 3** (a) 4D-STEM ptychography image displaying extra T atoms (indicated by arrows) relative to the typical MFI framework. (b) Structural projections of straight 10-MR channels of MFI, MEL, and their superposition. Symbols *i* and *m* denote inversion and mirror symmetry, respectively. (c, d) Second (c) and fourth (d) slices of the multislice ptychography reconstruction result. Raw images are provided in Fig. S12. The entire reconstructed volume is presented in Video S1 to display slice-by-slice structural changes. (e) Upper panel: enlarged image from the yellow rectangle region in (d), showing the interface of MFI and MEL within the pentasil chain. Middle and bottom panels: structural model of the interface viewed along different orientations, in which T atoms of MFI and MEL are all kept at the interface to demonstrate minimal mismatch. (f) Structural model of MFI/MEL intergrowth projected along the *c*-axis. (g) Outlined bar area in (f) projected along the *a*-axis, displaying the disconnected silanol-terminated interface perpendicular to the *b*-axis. In (f) and (g), red and gray balls represent O and hydrogen atoms, respectively. Green and orange balls represent T atoms in MFI and MEL, respectively.

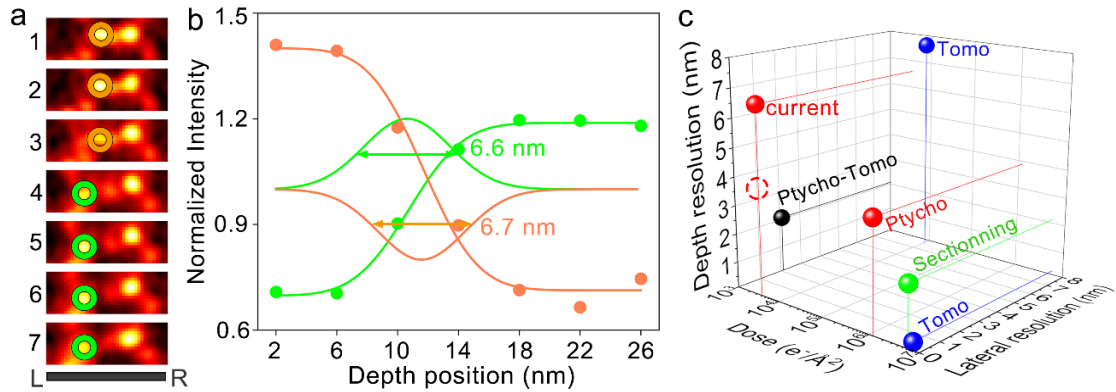

**Fig. 4.** (a) Slice-by-slice analysis of the region highlighted by the yellow rectangle in Fig. 3a. Each slice is 4 nm thick. Green and orange circles mark the T columns in MFI and MEL, respectively. (b) Depth resolution estimation based on intensity profiles. Normalized intensities at the two locations indicated by green and orange circles in (a) are extracted from seven successive slices and plotted as a function of their depth positions. The intensities were normalized by their mean values. The intensity variations at the two locations, represented by green and yellow dots, are fitted with the error function. The full width at half maximum of the derivative (green and yellow lines) of the fitted error function is defined as the depth resolution. (c) Lateral and depth resolution reported for various 3D electron microscopy imaging methods. Tomo, Sectioning, Ptycho, and Ptycho-Tomo denote tomography, high-angle annular dark-field STEM depth sectioning, ptychography, and ptychographic tomography, respectively. The dashed circle denotes the depth resolution of the ptychographic phase image reconstructed from a simulated 4D-STEM dataset.

Supplementary Materials

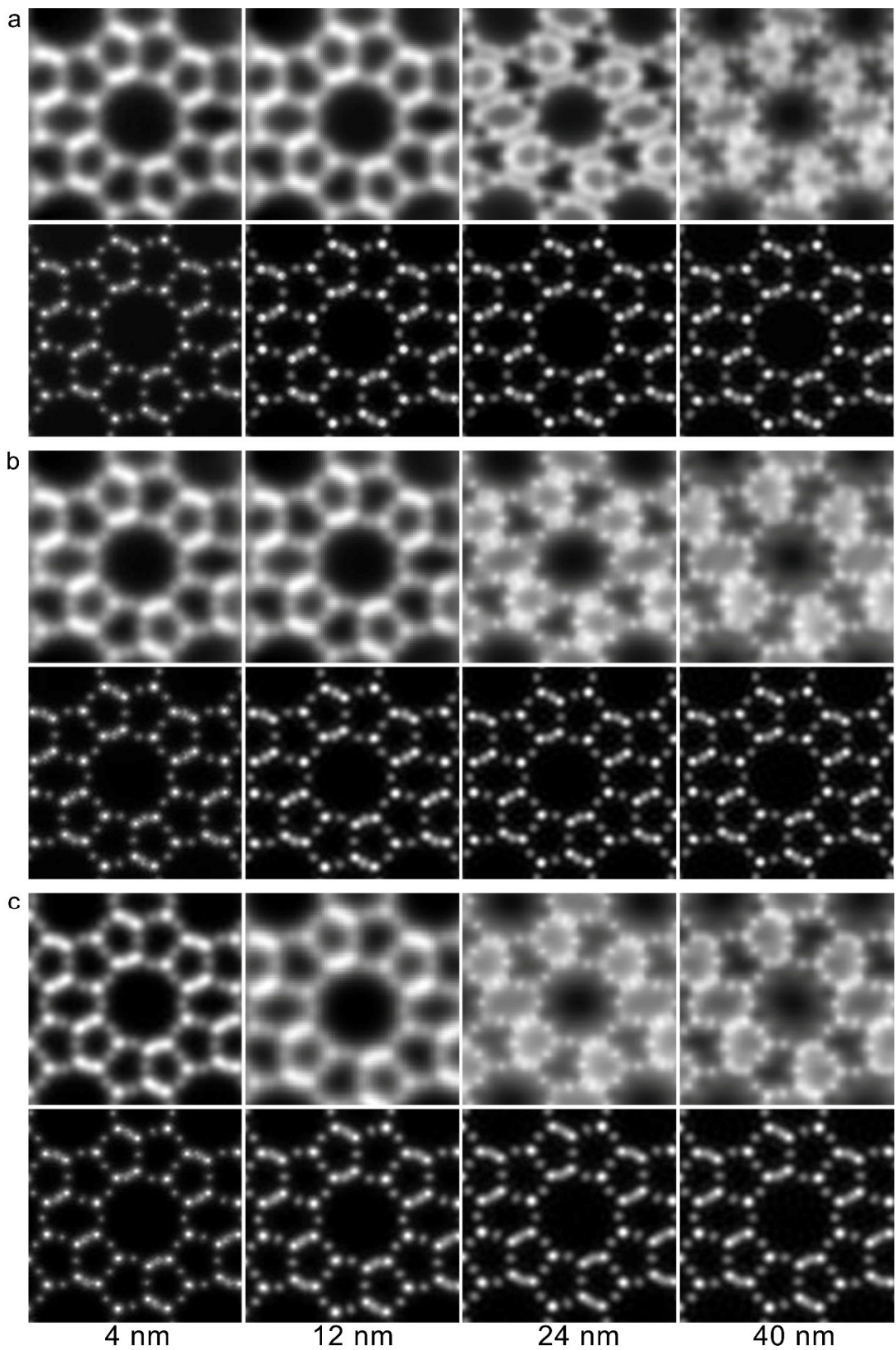

| | 4 nm | 12 nm | 24 nm | 40 nm |

Fig. S1. iDPC-STEM (first row in a, b, and c) and 4D-STEM ptychographic images (second row in a, b, and c) of [010]-projected ZSM-5 simulated using various specimen thicknesses and electron beam convergence angles: (a) 10 mrad, (b) 15 mrad, and (c) 21 mrad. Regardless of the convergence angle, the quality of iDPC images dramatically deteriorates as the specimen thickness exceeds 12 nm, whereas ptychographic phase images display excellent tolerance to thickness variation.

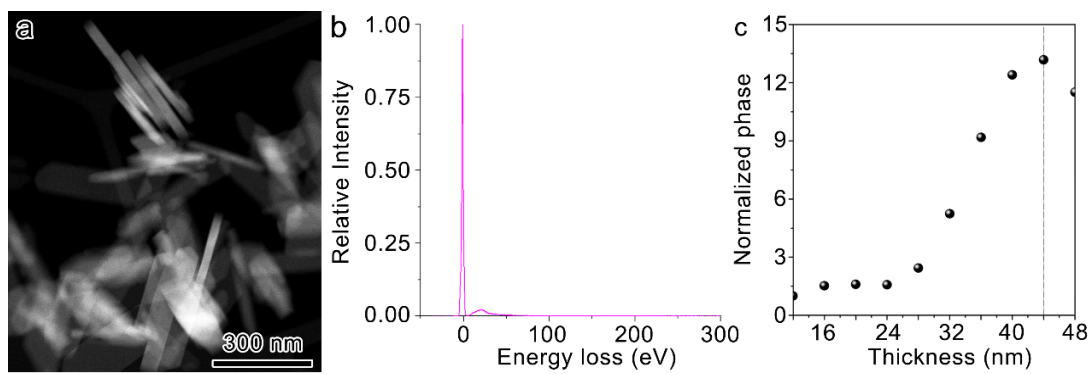

Fig. S2. (a) High-angle annular dark-field image, where several "standing" ZSM-5 crystals exhibit uniform thicknesses of 30–40 nm. (b) Electron energy loss spectrum collected from a "face-on" ZSM-5 crystal. The crystal thickness is estimated at ~47 nm based on the mean free path of inelastic scattering determined using David Mitchell's mean free path estimator script (http://www.dmscripting.com/meanfreepathestimator.html). (c) Thickness dependence of the ptychography-reconstructed image in Fig. 1b. Given that the reconstruction with a thickness close to the actual specimen thickness offers the maximum phase, the specimen thickness of Fig. 1b is ~44 nm. In (c), the phases are normalized by the value reconstructed with a 12 nm thickness.

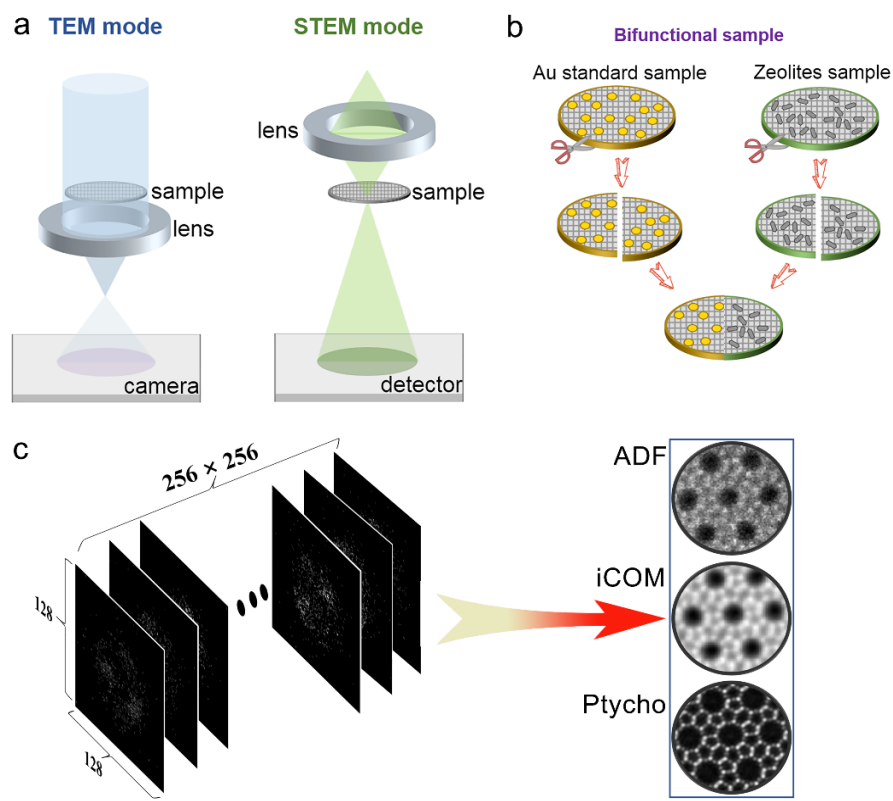

Fig. S3. (a) Schematic illustrations of TEM and STEM modes. (b) Strategy to minimize aberrations for 4D-STEM data acquisition. (c) 4D-STEM dataset consisting of 256 × 256 diffraction patterns recorded using an EMPAD detector with 128 × 128 pixels. Annular dark-field (ADF) and integrated center-of-mass (iCOM) images were used for the initial screening of 4D-STEM datasets, where iCOM is a precise analog of iDPC. Ptychographic phase (Ptycho) images were reconstructed using combined maximum likelihood and linear least squares methods with multislice and mixed-state algorithms.

Note: Due to the electron beam sensitivity of zeolites, the sample search and tilt were performed in TEM mode rather than STEM mode to avoid beam damage[1–3]. However, toggling between TEM and STEM modes deteriorates aberrations, especially $A_1$ and $B_2$. Although electron ptychography is, in theory, aberration-independent, the ptychographic reconstruction using low-dose data requires the initially guessed probe to be as close to the actual probe as possible. Therefore, aberrations except defocus should be corrected as much as possible.

To obtain optimal probing conditions, we prepared a special TEM grid composed of two half grids, supporting the Au standard sample for aberration correction and the zeolite sample to be studied, respectively. The workflow is as follows:

(i) Correct aberrations as much as possible using the Au standard in STEM mode at a high probe current (~150 pA).

(ii) Switch to TEM mode to search for the zeolite crystal of interest and align the zone axis at a magnification of 13,000 with a dose rate of ~0.3 e$^-$/Å$^2$/s.

(iii) Return to the Au-containing half grid, switch to STEM mode and recorrect aberrations ($A_1$, $B_2$) using the Ronchigram at a probe current of ~10 pA.

(iv) Carefully correct $A_1$ at a probe current of ~0.1 pA using Au particles at a few million magnifications and an extended dwell time (40~60 μs).

(v) Move to the recorded region of interest on the zeolite-containing half grid and stabilize the microscope for 2~4 h before collecting 4D-STEM data.

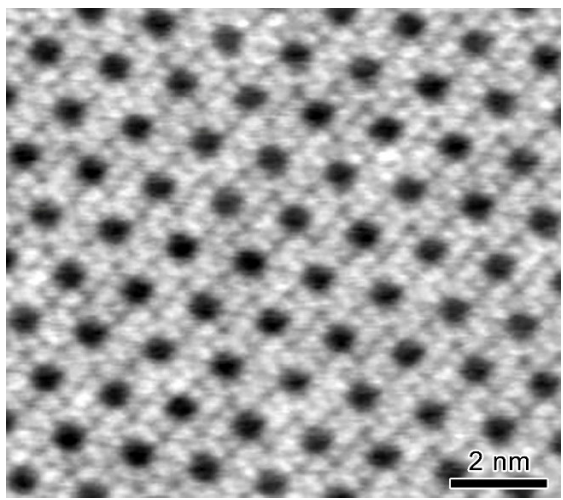

Fig. S4. iDPC image acquired along the [010] axis from the bulk region of an approximately 40-nm-thick ZSM-5 crystal.

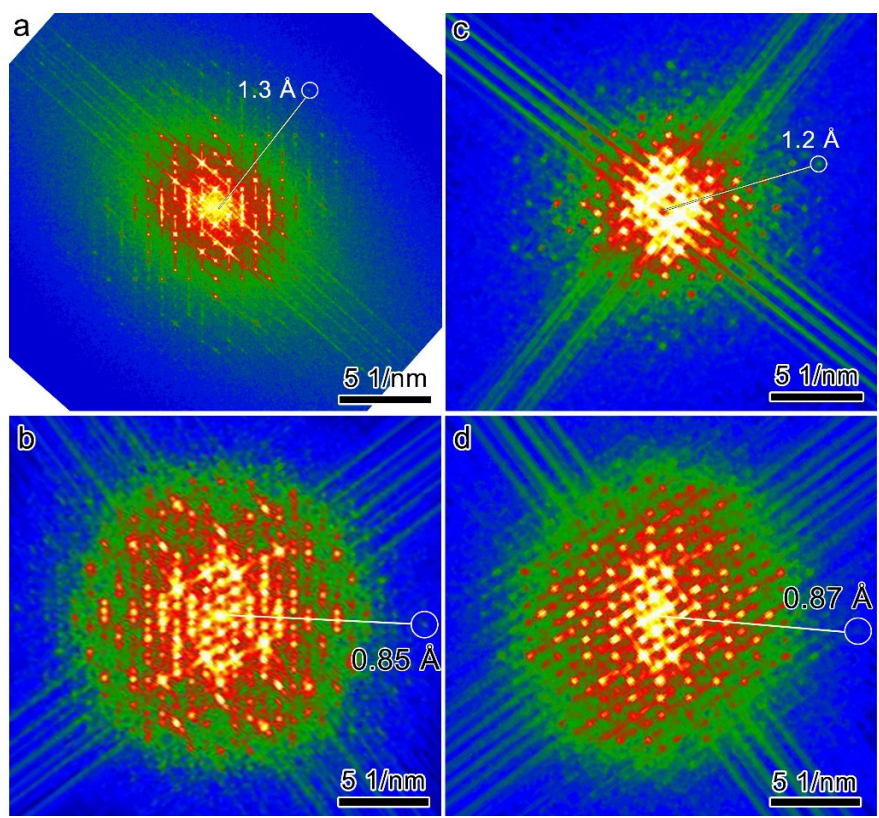

Fig. S5. (a-d) Fast Fourier transforms of the full images of Figs. 1a, 1b, 1d, and 1e.

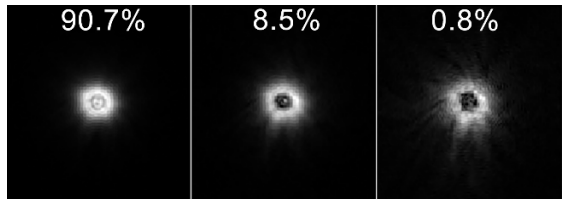

Fig. S6. The strongest three of the five orthogonal probe modes used in the multislice reconstruction of Fig. 1b, with their respective contributions to the total illumination indicated as a percentage. The defocus of the probe is estimated at ~35 nm.

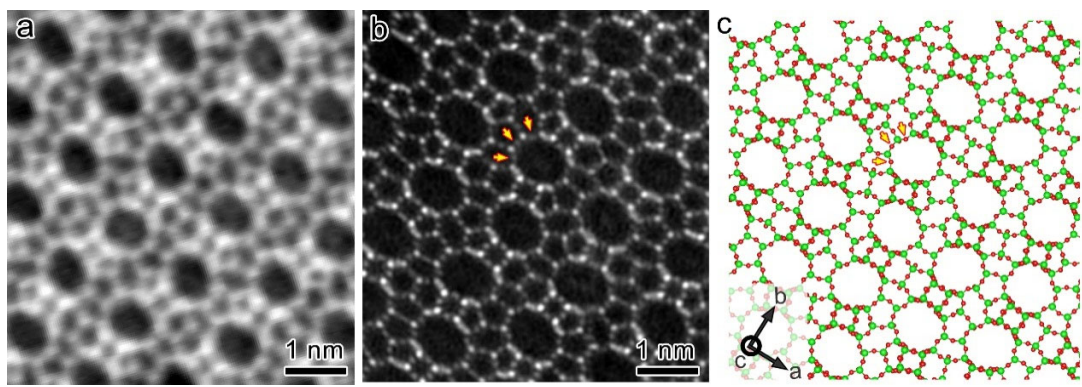

Fig. S7. (a) iDPC-STEM and (b) 4D-STEM ptychography image acquired along the characteristic 11-MR channel direction (i.e., the [001] axis) of EMM-17. (c) The [001]-projected structural model of EMM-17. Arrows in (b) and (c) indicate several O atoms (strictly, atomic columns) in the framework, which are not resolved in (a). The thickness of the imaged crystal is estimated at ~36 nm based on the ptychography reconstruction result.

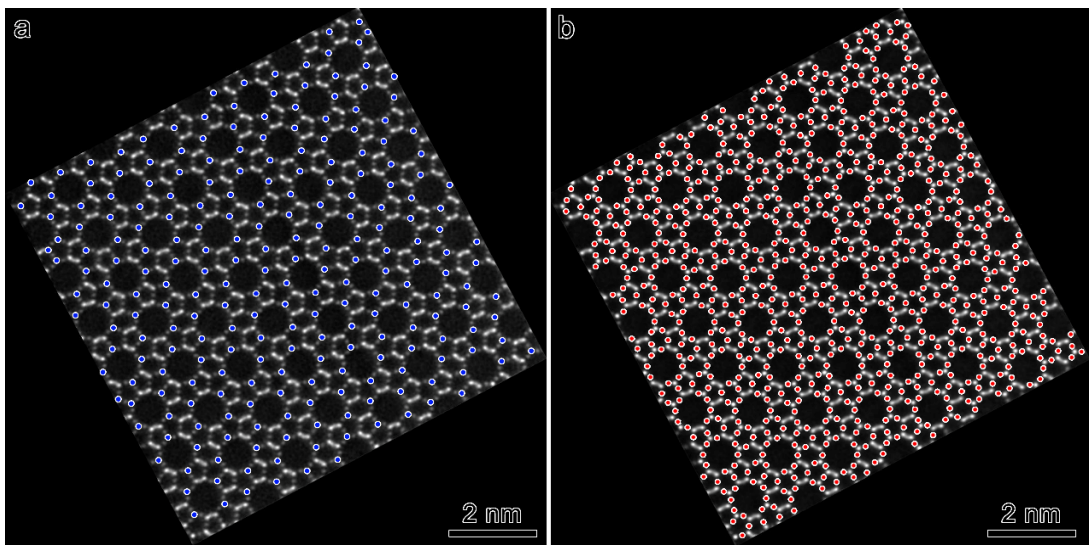

Fig. S8. Identification of T (a) and O (b) atomic columns from the ptychography-reconstructed images for subsequent intensity analysis. The atomic columns are identified, classified, and integrated using the Atomap package[4]. Partially overlapped columns were excluded from the analysis. This method was applied to the simulated and experimental images to generate the statistical results presented in Fig. 2a and 2b.

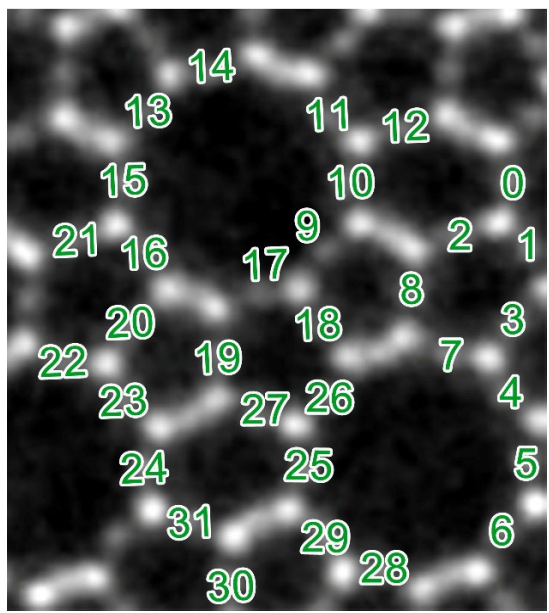

Fig. S9. The 30 O atomic columns in Fig. 2c illustrate how O vacancies were identified by comparison with the threshold determined in simulations. The image corresponds to one of the slices reconstructed from the experimental dataset of commercial ZSM-5. Atomic column identification and intensity analysis were performed using the Atomap package.

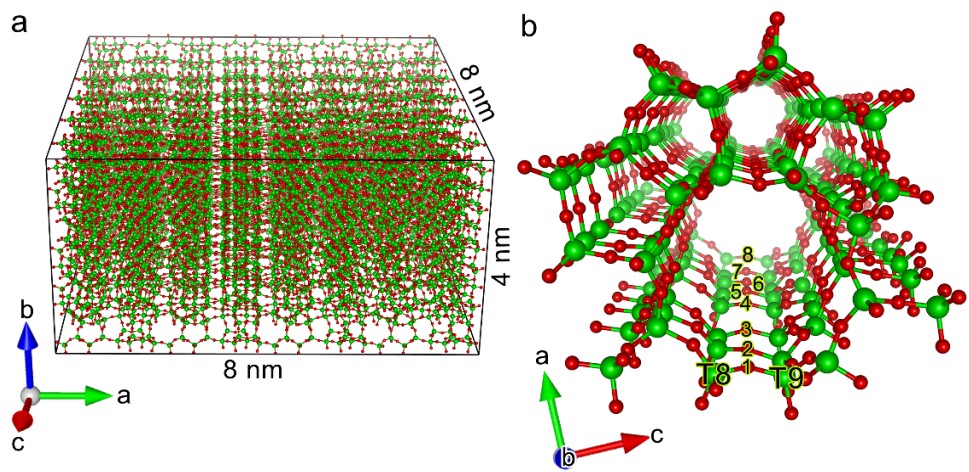

Fig. S10. (a) MFI structural model with dimensions of 8 nm (*a*-axis) × 8 nm (*c*-axis) × 4 nm (*b*-axis). Each oxygen column along the *b*-axis has eight oxygen atoms. (b) Illustration of where to introduce O vacancies for simulation.

Notes: One of the O columns between the T8 and T9 columns near the center of the structural model (a) is selected for introducing vacancies, and the eight O sites in the selected column are numbered (b).

When one O vacancy was introduced at eight different sites in the column for simulation, the obtained eight column intensities (relative to the strongest O column within the 1 nm range) determine the one-vacancy intensity range described in Fig. 2c. This range largely overlaps with the intrinsic O column intensity fluctuation of ZSM-5 (Fig. 2c), suggesting that one O vacancy within a thickness of 4 nm (i.e., one vacancy per eight O atoms) cannot be reliably determined.

Introducing two vacancies at eight different O sites results in 18 combinations, whose relative intensities obtained from simulations are summarized in Table S1. The lower limit of the two-vacancy intensity range was used as a threshold to identify the presence of O vacancies from the experimental ptychographic phase images (Fig. 2c).

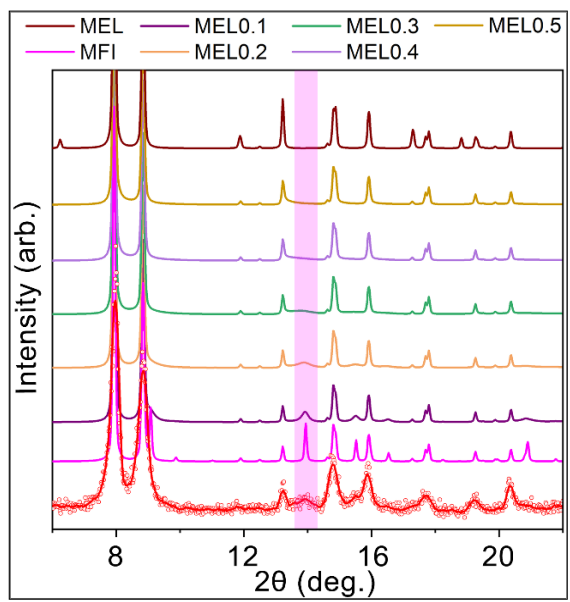

Fig. S11. PXRD patterns of MFI, MEL, and randomly stacked MFI/MEL with various MEL content were simulated using DIFFaX[5] compared to the experimental data (red dots). The comparison based on the marked peak intensity reveals that the MEL content is 10%−20%.

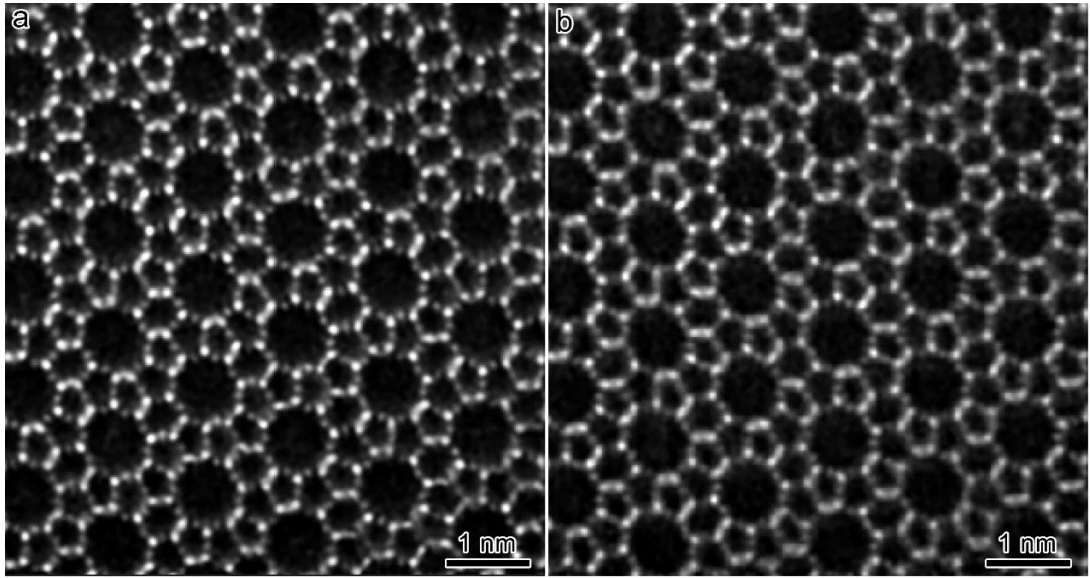

Fig. S12. Raw images of Fig. 3c (a) and Fig. 3d (b).

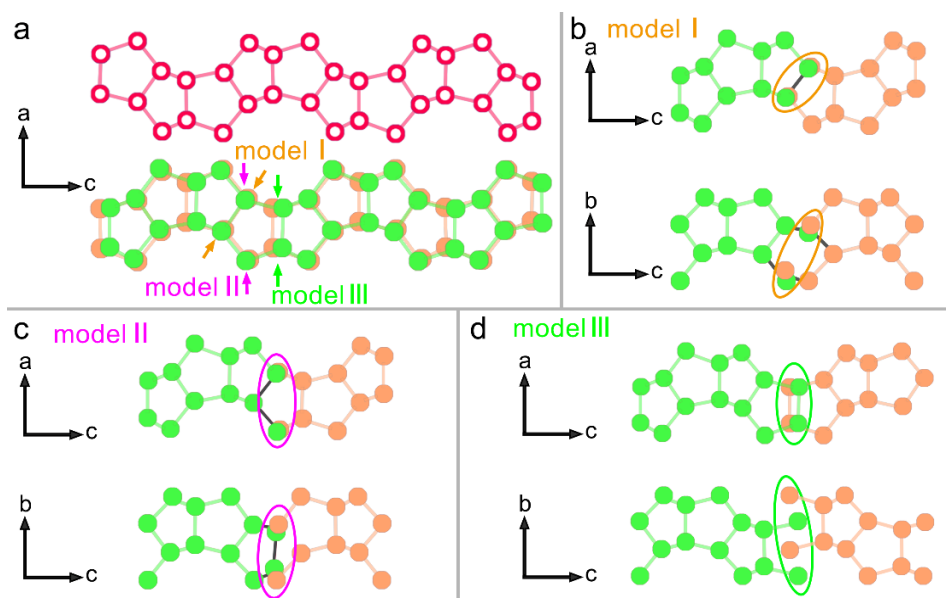

Fig. S13. (a) By connecting the upper red pentasil chain, the green and orange pentasil chains form MFI and MEL structures, respectively. The green and orange pentasil chains can be used to understand how MFI and MEL are connected within a pentasil chain. The three basic connection models are I, II, and III. (b-d) Models I, II, and III are viewed in different directions, respectively. In models I and II, the T sites of the MFI and MEL segments are well matched at the interface when viewed along the *b*-axis and *a*-axis. In model III, the T sites of the MFI and MEL segments cannot overlap, especially when viewed along the *a*-axis, suggesting an unfavorable disconnected interface. Model I was experimentally observed, as presented in Fig. 3e.

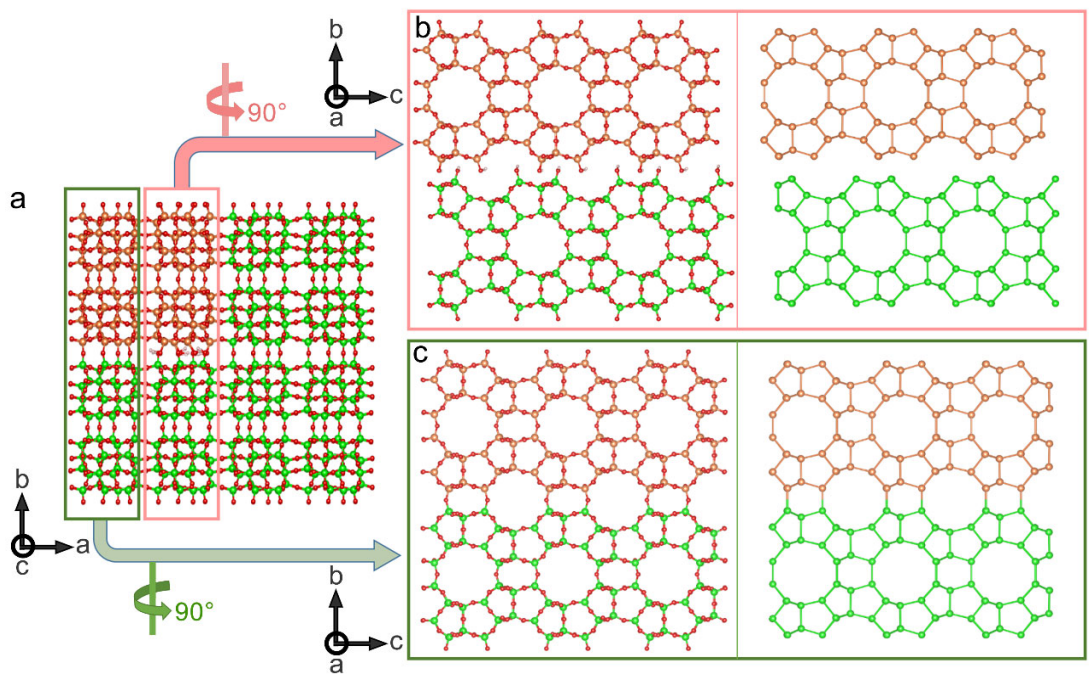

Fig. S14. (a) Structural model of MFI and MEL intergrowth, where T atoms in MFI and MEL are presented in orange and green, respectively. The MFI/MEL interface perpendicular to the *b*-axis has alternating "connected" and "disconnected" regions along the *a*-axis. (b-c) The disconnected (b) and connected (c) interfaces are viewed along the *a*-axis, where the left panels reveal all types of atoms, whereas the right panels display T atoms only. The disconnected silanol-terminated interface is clearly illustrated in (b).

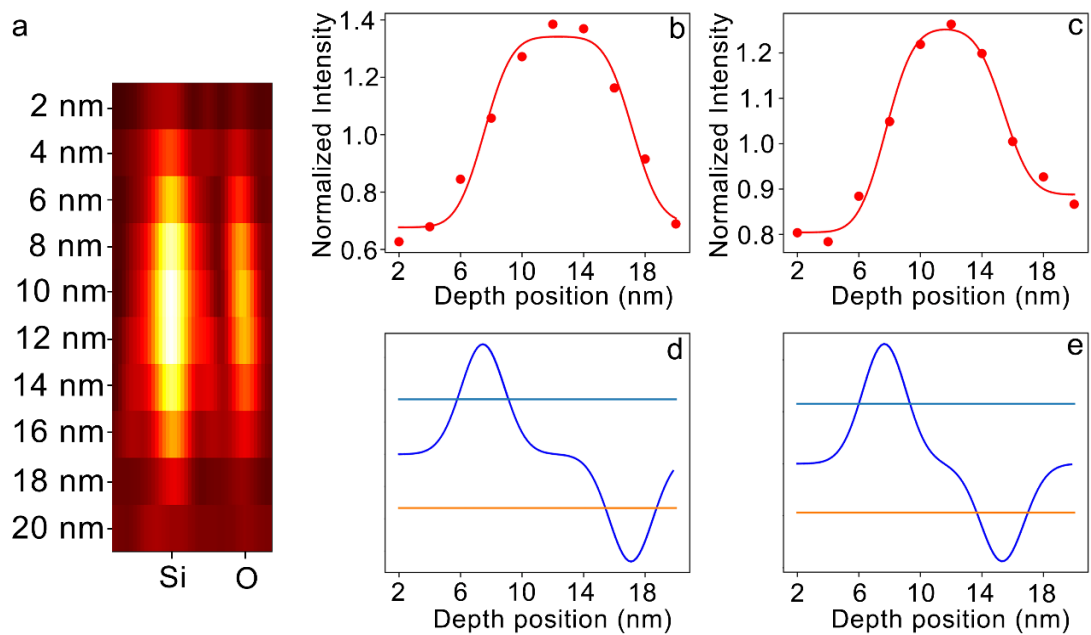

Fig. S15. (a) Cross-section of Si and O columns in the ptychography image reconstructed from a simulated 4D-STEM dataset of ZSM-5 at the dose level of ~3500 e⁻/Å². (b, c) The normalized intensities of Si (b) and O (c) were fitted with the error function. (d, e) Derivatives of the fitted curves in (b) and (c), respectively. The depth resolution determined from the full width at half maximum of the derivative peaks is approximately 3.5 nm.

**Table S1**. Intensities of the oxygen columns with two vacancies at different positions

|          | Vacancy position [a] | $I/I_{max}$ [b] |
|----------|----------------------|-----------------|
| **Model 1**  | 1&2 | 0.822 |
| **Model 2**  | 1&3 | 0.821 |
| **Model 3**  | 1&4 | 0.816 |
| **Model 4**  | 1&5 | 0.845 |
| **Model 5**  | 1&6 | 0.836 |
| **Model 6**  | 1&7 | 0.845 |
| **Model 7**  | 2&3 | 0.845 |
| **Model 8**  | 2&4 | 0.836 |
| **Model 9**  | 2&5 | 0.845 |
| **Model 10** | 2&6 | 0.838 |
| **Model 11** | 2&7 | 0.814 |
| **Model 12** | 3&4 | 0.826 |
| **Model 13** | 3&5 | 0.835 |
| **Model 14** | 3&6 | 0.835 |
| **Model 15** | 3&7 | 0.837 |
| **Model 16** | 4&5 | 0.819 |
| **Model 17** | 4&6 | 0.825 |
| **Model 18** | 4&7 | 0.826 |

[a] Numbers indicate the O vacancy sites in the model used for image simulation (e.g., "1&2" means that O sites 1 and 2 in Fig. S10b are vacant).

[b] I denotes the intensity of the designated vacancy-containing O column, and $I_{max}$ represents the maximum intensity of O columns in the 1 nm range from that column.

**Table S2.** Distribution of the O vacancy sites identified in Fig. 2d[a].

|     | R5&R5 | R5&R6 | R5&R10 | R6&R10 |
|-----|-------|-------|--------|--------|
| **L1** | 3     | NF    | 4      | NF     |
| **L2** | NF    | NF    | 2      | NF     |
| **L3** | NF    | NF    | 3      | NF     |
| **L4** | 1     | NF    | 4      | 1      |
| **L5** | 1     | NF    | 5      | 1      |
| **L6** | 4     | NF    | 2      | NF     |
| **L7** | 5     | 1     | 4      | 1      |

[a] R5, R6 and R10 represent the 5-, 6-, and 10-MR channels in the [010] projection. The O vacancy sites are predominantly located in R10, accounting for ~64%. NF: not found. Averaging the results of three datasets, approximately 58% of the identified O vacancies are exposed to 10-MR channels.

Table S3. Electron dose and resolution plotted in Fig. 4c

| Dose (e⁻/Å²) | Resolution (nm) | | References |
|---|---|---|---|
| | Lateral | Longitudinal | |
| $1.20 \times 10^6$ | 0.03 | 3.90 | 6 |
| 1035 | 2.08 | 2.08 | 7 |
| $5.64 \times 10^6$ * | 0.09 | 2.14 | 8 |
| $6.10 \times 10^6$ | 0.24 | 0.24 | 9 |
| $1.00 \times 10^4$ | 7.7 | 7.70 | 10 |
| 3500 | 0.085 | 6.60 | current |

\* The dose was obtained via private email communication with the authors

Video S1. Slice-by-slice view of the multislice ptychography reconstruction result of ZSM-5/ZSM-11 (MFI/MEL), revealing the longitudinal structural changes.